\documentclass{aipproc}
\layoutstyle{6x9}
\usepackage{bm}
\begin{document}

\title{Gluonic excitations in lattice QCD:\\ A brief survey}

\classification{12.38.Gc, 11.15.Ha, 12.39.Mk}
\keywords{Document processing, Class file writing, \LaTeXe{}}

\author{Colin Morningstar}{
  address={Department of Physics, Carnegie Mellon University, Pittsburgh,
           PA, USA  15213-3890},
  email={colin_morningstar@cmu.edu},
}

\copyrightyear  {2001}

\begin{abstract}
  Our current knowledge about glueballs and hybrid mesons from lattice
QCD simulations is briefly reviewed.
\end{abstract}

\date{\today}
\maketitle

\section{Introduction}

Hadronic states bound together by an {\em excited} gluon field, such
as glueballs, hybrid mesons, and hybrid baryons, are a potentially
rich source of information concerning the confining properties of QCD.
Interest in such states has been recently sparked by observations of
resonances with exotic $1^{-+}$ quantum numbers\cite{E852}
at Brookhaven.  In fact,
the proposed Hall D at Jefferson Lab will be dedicated to the search
for hybrid mesons and one of the goals of CLEO-c will be to identify
glueballs and exotics.  Although our understanding of these states
remains deplorable, recent lattice simulations have shed some light
on their nature.  In this talk, I summarize our current knowledge
about glueballs and heavy- and light-quark hybrid mesons from lattice
QCD simulations.

\section{Glueballs}

\begin{figure}[t]
\includegraphics[width=10cm,bb= 29 86 552 482 ]{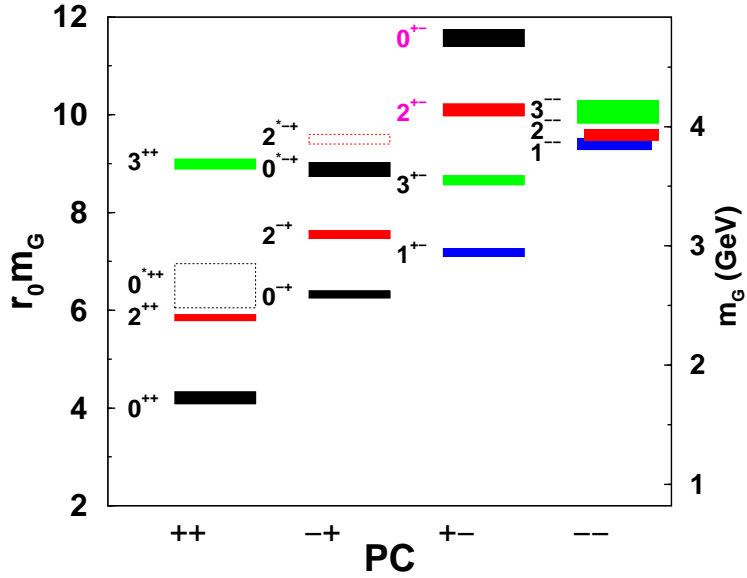}
\caption{ The mass spectrum of glueballs in the pure SU(3) gauge theory
 from Ref.~\protect\cite{glueballs}.
 The masses are given in terms of the hadronic scale $r_0$ along the
 left vertical axis and in terms of GeV along the right vertical axis
 (assuming $r_0^{-1}=410(20)$ MeV).  The mass uncertainties indicated by the
 vertical extents of the boxes do {\em not} include the uncertainty in
 setting $r_0$.  The locations of states whose interpretation requires
 further study are indicated by the dashed hollow boxes.
\label{fig:contglue}}
\end{figure}

The glueball spectrum in the absence of virtual quark-antiquark
pairs is now well known\cite{glueballs} and is shown in
Fig.~\ref{fig:contglue}.  The use of spatially coarse, 
temporally fine lattices with improved actions and the application
of variational techniques to moderately-large matrices of
correlations functions were crucial in obtaining this spectrum.
Continuum limit results for the lowest-lying scalar and tensor states
are in good agreement with recent previous calculations\cite{bali,gf11}
when expressed in terms of the hadronic scale $r_0$, the square
root of the string tension $\sqrt{\sigma}$, or in terms of each other
as a ratio of masses.  Disagreements in specifying the masses of these
states in GeV arise solely from ambiguities in setting the value
of $r_0$ within the quenched approximation.  The glueball spectrum
can be qualitatively understood in terms of the interpolating
operators of minimal dimension which can create glueball 
states\cite{jaffe} and can be reasonably well explained\cite{kuti}
in terms of a simple constituent gluon (bag) model which approximates
the gluon field using spherical cavity Hartree modes with residual 
perturbative interactions\cite{gluebag1,gluebag2}.  The challenge
now is to deduce precisely what the spectrum in Fig.~\ref{fig:contglue}
is telling us about the long-wavelength properties of QCD.  This
spectrum provides an important testing ground for models of confined
gluons, such as center and Abelian dominance, instantons, soliton
knots, instantaneous Coulomb-gauge mechanisms, and so on.

\begin{figure}[b]
\includegraphics[width=8cm,bb=48 26 507 367]{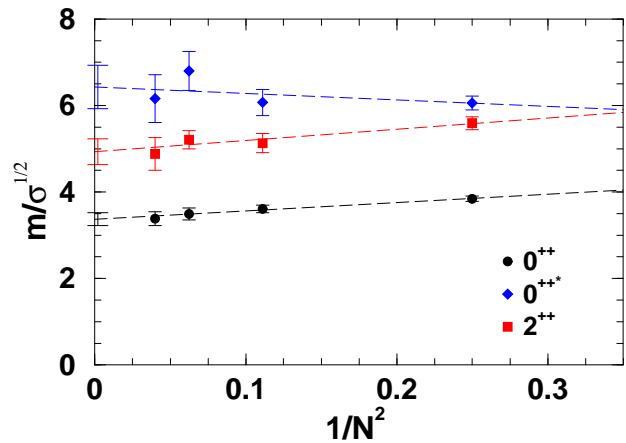}
\caption{Continuum scalar, tensor, and excited scalar $SU(N)$ glueball
masses expressed in units of the string tension $\sigma$ and plotted
against $1/N^2$. Linear extrapolations to $N=\infty$
are shown in each case (see Ref.~\protect\cite{teper}). } 
\label{fig_glueN}
\end{figure}

Recently, the two lowest-lying scalar glueballs and the tensor glueball
have been studied\cite{teper} in $SU(N)$ for $N=2,3,4,$ and $5$.  Their
masses have been shown to depend linearly on $1/N^2$, and these masses
in the limit $N\rightarrow\infty$ do not differ substantially from
their values for small $N$ (see Fig.~\ref{fig_glueN}).

Glueball wavefunctions and sizes have been
studied in the past, but much of the early work contains uncontrolled
systematic errors, most notably from discretization effects.
The scalar glueball is particularly susceptible to such errors
for the Wilson gauge action due to the presence of a critical end point
of a line of phase transitions (not corresponding to any physical
transition found in QCD) in the fundamental-adjoint coupling 
plane.  As this critical point (which defines the continuum limit
of a $\phi^4$ scalar field theory) is neared, the coherence length
in the scalar channel becomes large, which means that the mass gap in
this channel becomes small; glueballs in other channels seem
to be affected very little.  Results in which the scalar glueball
was found to be significantly smaller than the tensor were most
likely due to contamination of the scalar glueball from the
non-QCD critical point.  In Ref.~\cite{gluewavefunc}, an improved
gauge action designed to avoid the spurious critical point found that
the scalar and tensor glueballs were comparable in size and of typical
hadronic dimensions.  Operator overlaps obtained from the variational
optimizations carried out in Ref.~\cite{glueballs}, which also used
an improved gauge action, concur with such a conclusion.  

Vacuum-to-glueball transition amplitudes for operators
such as $\langle G\vert {\rm Tr}\bm{E}^2\vert 0\rangle$ and
$\langle G\vert {\rm Tr}\bm{B}^2\vert 0\rangle$ in the $0^{++}$ sector,
$\langle G\vert {\rm Tr}\bm{E\cdot B}\vert 0\rangle$ in the $0^{-+}$
sector, and $\langle G\vert {\rm Tr}\bm{E}_i\bm{E}_j\vert 0\rangle$
in the $2^{++}$ sector, have also been studied in the past\cite{gluematrix}.
Efforts to revisit these calculations using improved actions and operators
with reduced discretization errors is ongoing.

\begin{figure}
\includegraphics[width=11cm,bb=12 22 576 262]{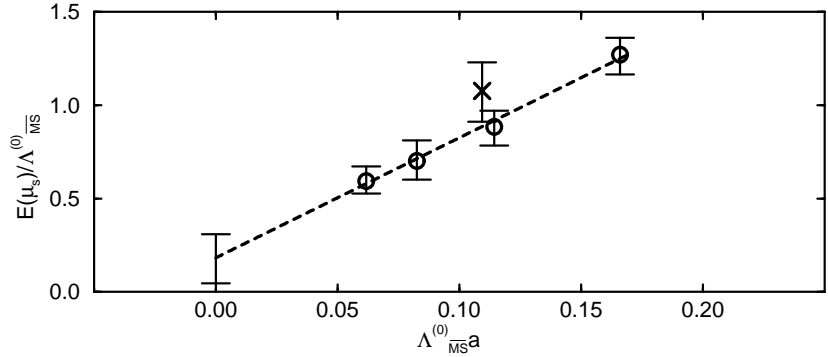}
\caption{Lattice spacing dependence and continuum limit of the
glueball-quarkonium mixing energy $E(\mu_s)$ at the strange quark
mass in the quenched approximation in terms of 
$\Lambda^{(0)}_{\overline{MS}}=234.9(6.2)$ MeV set using the
(quenched) mass of the $\rho$ meson. The continuum limit value 
of $E(\mu_s)$ is 43(31) MeV. The circles indicate results using
lattices with spatial extent approximately 1.6 fm; the $\times$ indicates
a result in a larger volume with spatial extent near 2.3 fm.
(see Ref.~\protect\cite{gluemix}).} 
\label{fig_glue_mix}
\end{figure}

A valiant effort to study the decay of the scalar glueball into two
pseudoscalar mesons was made in Ref.~\cite{gluedecay}.  A slight
mass dependence of the coupling was found, and a total width of
$108(29)$ MeV for decays to two pseudoscalars was obtained in
the quenched approximation.  However, these results were obtained
with the Wilson gauge action at $\beta=5.7$, an unfortunate choice
since pollution of the scalar glueball from the non-QCD critical point
is significant (for example, the mass differs from its continuum
limit value by at least 20\% and the mixing energy of the glueball
with quarkonium differs dramatically from its continuum limit value
as discussed below).  These large systematic
uncertainties and the use of the quenched approximation should be
kept in mind when considering the value of the width given above.

Mixing of the scalar glueball with quarkonium has been studied
in the quenched approximation in Ref.~\cite{gluemix}.  Several
lattice spacings were used to facilitate control of discretization
errors.  Good control of systematic errors was demonstrated, although
extrapolations in the quark mass to $\mu_n$ 
(the $u$ and $d$ quark mass scale) might benefit from
the use of chiral perturbation theory.  The results for the mixing
energy are shown in Fig.~\ref{fig_glue_mix}.  Note that the
continuum limit is essentially {\em consistent with zero mixing}. 
Again, remember that the inclusion of quark loops is incomplete when
assessing this conclusion.  The
large variation of this mixing energy with the lattice spacing is
most likely due to the use of the simple Wilson gauge action.  The
rightmost point corresponds to $\beta=5.7$ which illustrates the very
large lattice artifacts at this spacing.  This mixing calculation
could benefit enormously from the use of an improved action and
anisotropic lattices.

\begin{figure}
\includegraphics[width=10cm,bb= 50 50 554 482]{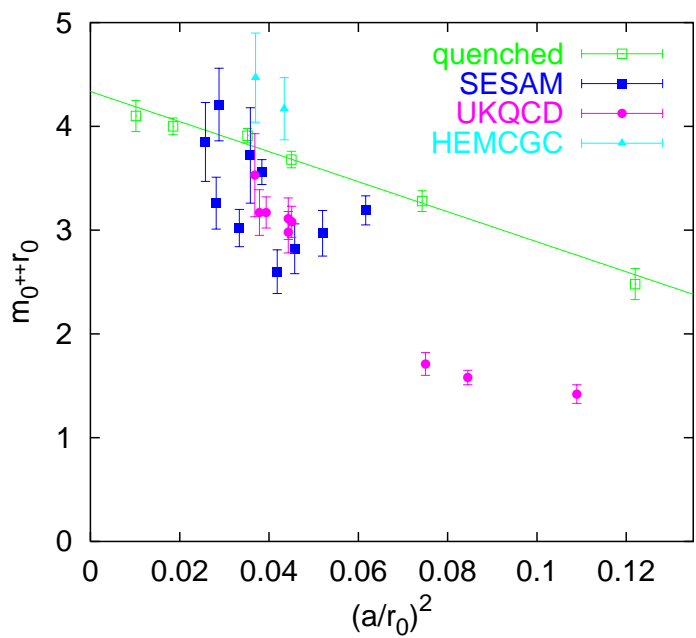}
\caption{ The scalar glueball mass with two flavors of virtual
 quark-antiquark pairs included against the lattice spacing $a$ in
 terms of the hadronic scale $r_0$, courtesy of 
 Ref.~\protect\cite{gluefull}.  Results are shown for
 three different discretizations of the Dirac action: staggered
 (HEMCGC, from Ref.~\protect\cite{HEMCGC}),
 Wilson (SESAM from Ref.~\protect\cite{SESAM}),
 and clover (UKQCD from Refs.~\protect\cite{UKQCD1,UKQCD2}).
 The quarks are all heavier than a third of the strange quark
 mass. Quenched results are shown for comparison.
\label{fig:gluefull}}
\end{figure}

Incorporation of virtual quark-antiquark pairs in calculating the
glueball/quarkonium spectrum is a daunting task.  The fermion determinant
must be included in the Monte Carlo updating, dramatically increasing
the computational costs.  Mixings with two meson states require
all-to-all propagators, further adding to the cost and the
stochastic uncertainties.  Instabilities of the higher lying
states must be properly taken into account (such as with finite
volume techniques).  Extrapolations to realistically light quark
masses must be done carefully, taking decay thresholds into
account and possibly requiring simulations to be carried out at
quark masses lighter than currently feasible. Nevertheless, a few
groups\cite{HEMCGC,SESAM,UKQCD1,UKQCD2} have begun glueball/meson
simulations with two flavors of sea quarks, and the results for the
scalar glueball mass are shown in Fig.~\ref{fig:gluefull}.  The
status of such calculations has recently been reviewed in
Ref.~\cite{gluefull}.  The quarks are still heavier than $m_s/3$, where
$m_s$ is the mass of the strange quark.  The glueball mass tends to
decrease as the light quark mass is reduced, but it increases as
the lattice spacing is reduced, making it completely unclear what the
end result will be in the continuum limit for realistically light
quark masses.

\section{Heavy-quark hybrid mesons}

\begin{figure}
\includegraphics[width=12cm, bb= 167 344 488 582]{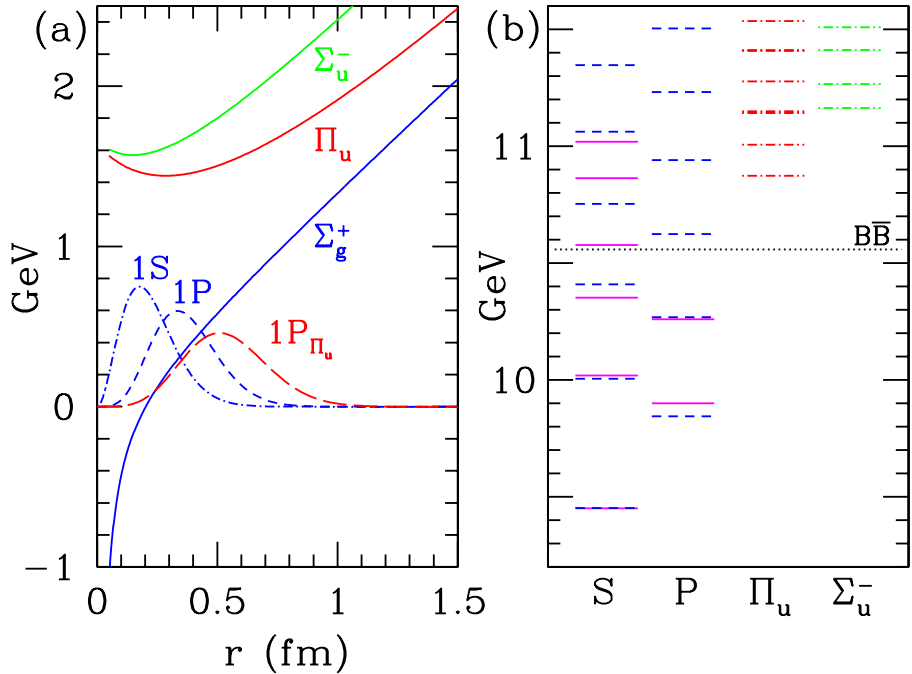}
\caption{(a) Static potentials and radial probability densities
 against quark-antiquark separation $r$ for $r_0^{-1}=450$ MeV. 
 (b) Spin-averaged $\bar{b}b$ spectrum
 in the LBO approximation (light quarks neglected).  Solid lines
 indicate experimental measurements.  Short dashed lines indicate 
 the $S$ and $P$ state masses obtained using the $\Sigma_g^+$ potential
 with $M_b=4.58$ GeV. Dashed-dotted
 lines indicate the hybrid quarkonium states obtained from the $\Pi_u$
 $(L=1,2,3)$ and $\Sigma_u^-$ $(L=0,1,2)$ potentials.  These results
 are from Ref.~\protect\cite{jkm}.
\label{fig:LBO}}
\end{figure}

One expects that a heavy-quark meson can be treated similar to a diatomic
molecule: the slow valence heavy quarks correspond to the nuclei and the fast
gluon and light sea quark fields correspond to the electrons\cite{hasenfratz}.
First, the quark $Q$ and antiquark $\overline{Q}$ are treated as
static color sources and the energy levels of the fast degrees of freedom
are determined as a function of the $Q\overline{Q}$ separation $r$,  each such
energy level defining an adiabatic surface or potential.  The motion of the
slow heavy quarks is then described in the leading Born-Oppenheimer (LBO)
approximation by the Schr\"odinger equation using each of these potentials.
Conventional quarkonia are based on the lowest-lying potential; hybrid
quarkonium states emerge from the excited potentials.

The spectrum of the fast gluon field in the presence of a static
quark-antiquark pair has been determined in lattice 
studies\cite{earlier1,earlier2}.
The three lowest-lying levels are shown in Fig.~\ref{fig:LBO}.  Due
to computational limitations, sea quark effects have been neglected in
these calculations; their expected impact on the hybrid meson spectrum
will be discussed below.  The levels in Fig.~\ref{fig:LBO} are labeled
by the magnitude $\Lambda$ of the projection of the total angular momentum
${\bf J}_g$ of the gluon field onto the molecular axis, and by $\eta=\pm 1$,
the symmetry under the charge conjugation combined with spatial inversion
about the midpoint between the $Q$ and $\overline{Q}$.  States with
$\Lambda=0,1,2,\dots$ are denoted by $\Sigma, \Pi, \Delta, \dots$,
respectively.  States which are even (odd) under the above-mentioned
$CP$ operation are denoted by the subscripts $g$ ($u$).  An additional
$\pm$ superscript for the $\Sigma$ states refers to even or odd symmetry
under a reflection in a plane containing the molecular axis.  The
potentials are calculated in terms of the hadronic scale parameter
$r_0$; in Fig.~\ref{fig:LBO}, $r_0^{-1}=450$ MeV has been assumed. 

The LBO spectrum\cite{jkm} of conventional $\overline{b}b$
and hybrid $\overline{b}gb$ states are shown in Fig.~\ref{fig:LBO}. 
Below the $\overline{B}B$ threshold, the LBO results are in very good
agreement with the spin-averaged experimental measurements of bottomonium
states.   Above the threshold, agreement with experiment is lost, suggesting
significant corrections either from mixing and other higher-order effects or
(more likely) from light sea quark effects.  Note from the radial probability
densities shown in Fig.~\ref{fig:LBO} that the size of the hybrid state is
large in comparison with the conventional $1S$ and $1P$ states. 

\begin{figure}
\includegraphics[width=10cm,bb= 28 144 572 552]{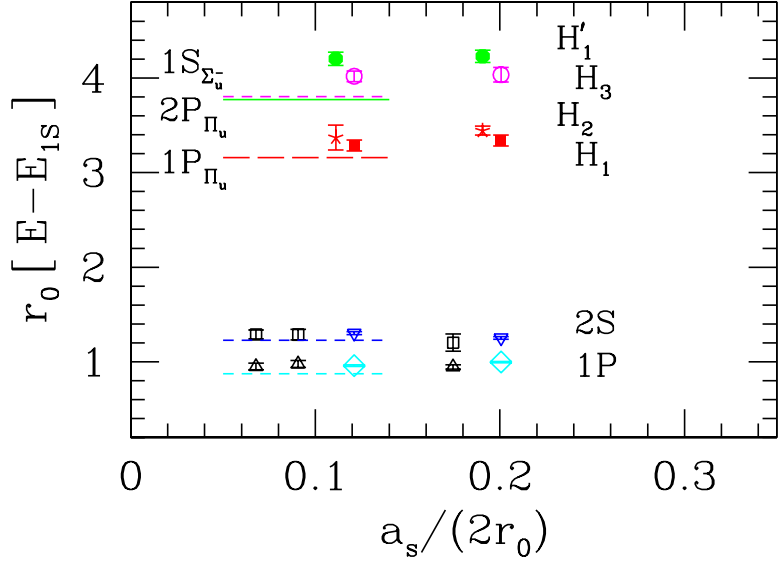}
\caption{
  Simulation results from Ref.~\protect\cite{jkm} for the heavy quarkonium
  level splittings
  (in terms of $r_0$ and with respect to the $1S$ state) against the lattice
  spacing $a_s$. Results from Ref.~\protect\cite{Wilson} using an NRQCD
  action with higher-order corrections are shown as open boxes and
  $\bigtriangleup$.  The horizontal lines show the LBO predictions.
  Agreement of these splittings within 10\% validates the Born-Oppenheimer
  approximation.
\label{fig:scaling}}
\end{figure}

The validity of such a simple physical picture relies on the smallness
of higher-order spin, relativistic, and retardation effects and mixings
between states based on different adiabatic surfaces.  
The importance of retardation and leading-order mixings between states
based on different adiabatic potentials can be tested by comparing the LBO
level splittings with those determined from meson simulations using a
leading-order non-relativistic (NRQCD) heavy-quark action.  Such a test was
carried out in Ref.~\cite{jkm}.  The NRQCD action included only a covariant
temporal derivative and the leading kinetic energy operator (with two other
operators to remove lattice spacing errors).  The only difference between the
leading Born-Oppenheimer Hamiltonian and the lowest-order NRQCD Hamiltonian
was the $\bm{p\!\cdot\!A}$ coupling between the quark color
charge in motion and the gluon field.   The level splittings (in terms of
$r_0$ and with respect to the $1S$ state) of the conventional $2S$ and $1P$
states and four hybrid states were compared (see Fig.~\ref{fig:scaling})
and found to agree within $10\%$, strongly supporting the validity of the
leading Born-Oppenheimer picture. 

The question of whether or not quark spin interactions spoil the 
validity of the Born-Oppenheimer picture for heavy-quark hybrids
has been addressed in Ref.~\cite{drummond}.  Simulations of several
hybrid mesons using an NRQCD action including the spin interaction
$c_B \bm{\sigma\!\cdot\! B}/2M_b$ and neglecting
light sea quark effects were carried out; the introduction
of the heavy-quark spin was shown to lead to significant level shifts
(of order 100 MeV or so) but the authors of Ref.~\cite{drummond} argue
that these splittings do {\em not} signal a breakdown of the 
Born-Oppenheimer picture.
First, they claim that no significant mixing of their non-exotic $0^{-+}$,
$1^{--}$, and $2^{-+}$ hybrid meson operators with conventional states was
observed; unfortunately, this claim is not convincing since a correlation
matrix analysis was not used.  Secondly, the authors argue that
calculations using the bag model support their suggestion.  These facts
are not conclusive evidence that heavy-quark
spin effects do not spoil the Born-Oppenheimer picture, but they are highly
suggestive. Further evidence to support the Born-Oppenheimer picture
has recently emerged in Ref.~\cite{toussaint}.  The NRQCD simulations
carried out in this work examined the mixing of the $\Upsilon$ with
a hybrid and found a very small probability admixture of hybrid in the
$\Upsilon$ given by $0.0035(1)c_B^2$ where $c_B^2\sim 1.5-3$ is expected.

\begin{figure}
\includegraphics[width=8cm,bb= 50 50 554 554 ]{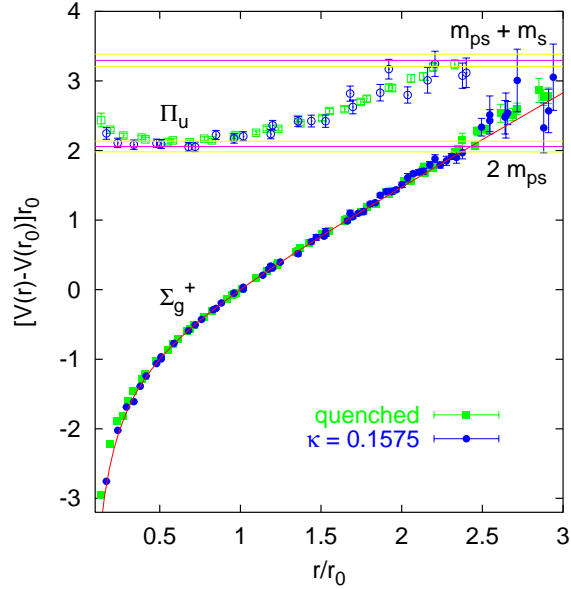}
\caption{ Ground $\Sigma_g^+$ and first-excited $\Pi_u$
 static quark potentials without sea quarks (squares, quenched) and
 with two flavors of sea quarks, slightly lighter than the strange
 quark (circles, $\kappa=0.1575$).  Results are given in terms of
 the scale $r_0\approx 0.5$ fm, and the lattice spacing is 
 $a\approx 0.08$ fm. Note that $m_S$ and $m_{PS}$ are the masses
 of a scalar and pseudoscalar meson, respectively,
 consisting of a light quark and a static antiquark.
 These results are from Ref.~\protect\cite{SESAM}.
\label{fig:hybridpot}}
\end{figure}

\begin{figure}
\includegraphics[width=8cm,bb= 73 48 507 402  ]{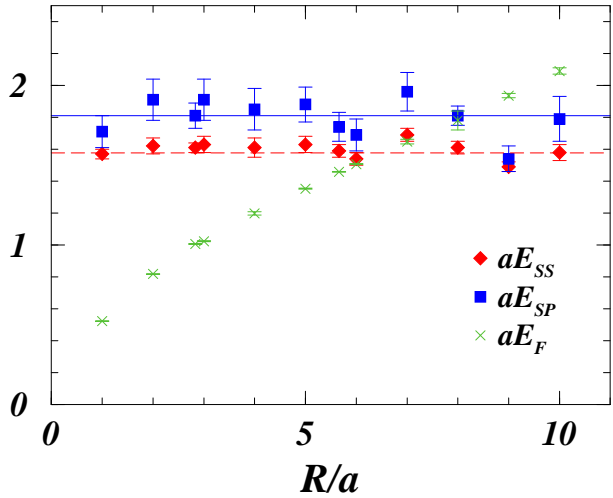}
\caption{ Evidence for ``string breaking'' at quark-antiquark
 separations $R\approx 1$ fm.  $E_{SS}$ is the energy of two
 $S$-wave static-light mesons (the light quark bound in an $S$-wave
 to the fixed static antiquark), $E_{SP}$ is the energy of an
 $S$-wave and a $P$-wave static-light meson, and $E_F$ is the
 energy of a static quark-antiquark pair connected by a gluonic
 flux tube.  The distance of separation $R$ refers to the distance
 between the static quark-antiquark pair.  All quantities are
 measured in terms of the lattice spacing $a\approx 0.16$ fm.
 Two flavors of light sea quarks are present with masses such
 that $m_\pi/m_\rho\approx 0.36$. The dashed and solid lines give
 the asymptotic values $2am_S$ and $a(m_P + m_S)$, where $m_S$ and
 $m_P$ are the masses of individual $S$-wave and $P$-wave
 static-light mesons, respectively.  Mixing between the flux tube
 and meson-meson channels was found to be very weak. Results are from 
 Ref.~\protect\cite{stringbreak}.
\label{fig:stringbreak}}
\end{figure}

The dense spectrum of hybrid states shown in Fig.~\ref{fig:LBO} neglects
the effects of light sea quark-antiquark pairs.  In order to include
these effects in the LBO, the adiabatic potentials must be determined
fully incorporating the light quark loops.  Such computations using
lattice simulations are very challenging, but good progress is being
made.  For separations below 1 fm, the $\Sigma_g^+$ and $\Pi_u$
potentials change very little\cite{SESAM} upon inclusion of the sea
quarks (see Fig.~\ref{fig:hybridpot}), suggesting that a few of the
lowest-lying hybrid states may exist as well-defined resonances. 
However, for $Q\overline{Q}$ separations greater than 1~fm, the
adiabatic surfaces change dramatically, as shown in 
Fig.~\ref{fig:stringbreak} from Ref.~\cite{stringbreak}.
Instead of increasing indefinitely, the static potential abruptly
levels off at a separation of 1 fm when the static quark-antiquark
pair, joined by flux tube, undergoes fission into two separate
$\overline{Q}q$ color singlets, where $q$ is a light quark.  Clearly, such
potentials cannot support the plethora of conventional and hybrid states
shown in Fig.~\ref{fig:LBO}; the formation of bound states and resonances
substantially extending over 1~fm seems unlikely.  Whether or not the
light sea quark-antiquark pairs spoil the Born-Oppenheimer picture is
currently unknown.  Future unquenched simulations should help to answer
this question, but it is not unreasonable to speculate that the simple
physical picture provided by the Born-Oppenheimer expansion for both
the low-lying conventional and hybrid heavy-quark mesons will survive
the introduction of the light sea quark effects.  Note that the
discrepancies of the spin-averaged LBO predictions with experiment
above the $B\overline{B}$ threshold seen in Fig.~\ref{fig:LBO} most
likely arise from the neglect of light sea quark-antiquark pairs.

\section{Light-quark hybrid mesons}
A summary of recent light-quark and charmonium $1^{-+}$ hybrid mass
calculations is presented in Table~\ref{table:light}.  With the exception
of Ref.~\cite{lasch98}, all results neglect light sea quark loops.  The
introduction of two flavors of dynamical quarks in Ref.~\cite{lasch98}
yielded little change to the hybrid mass, but this finding should not be
considered definitive due to uncontrolled systematics (unphysically large
quark masses, inadequate treatment of resonance properties in finite volume,
{\it etc.}).  All estimates of the light quark hybrid mass are near 
2.0 GeV, well above the experimental candidates found in the range
1.4-1.6 GeV.  Perhaps sea quark effects will resolve this discrepancy,
or perhaps the observed states are {\em not} hybrids.  Some authors have
suggested that they may be four quark $\bar{q}\bar{q}qq$ states.  Clearly,
there is still much to be learned about these exotic QCD resonances.

\begin{table}[t]
\begin{tabular}{llcl@{\hspace{6mm}}|@{\hspace{6mm}}lll}\hline
 \tablehead{4}{c}{b}{Light quark $1^{-+}$}
 &\tablehead{3}{c}{b}{Charmonium $1^{-+}\ -\ 1S$}\\[-2mm]
 \tablehead{2}{c}{b}{Ref.~\& Method} & 
 \tablehead{1}{c}{b}{$N_f$} &\tablehead{1}{l}{b}{$M$ (GeV)} &
 \tablehead{2}{c}{b}{Ref.~\& Method}
  & \tablehead{1}{c}{b}{$\Delta M$ (GeV)} \\ \hline
UKQCD 97\cite{ukqcd97}&SW&0&1.87(20)   &MILC 97\cite{milc97}     &W  &1.34(8)(20)\\
MILC 97\cite{milc97}  &W &0&1.97(9)(30)&MILC 99\cite{milc98}     &SW &1.22(15)   \\
MILC 99\cite{milc98}  &SW&0&2.11(10)   &CP-PACS 99\cite{cppacs99}&NR &1.323(13)  \\
LaSch 99\cite{lasch98}&W &2&1.9(2)     &JKM 99\cite{jkm}         &LBO&1.19       \\ \hline
\end{tabular}
\caption{
  Recent results for the light quark and charmonium $1^{-+}$ hybrid meson
  masses.  Method abbreviations: W = Wilson fermion action; SW = improved
  clover fermion action; NR = nonrelativistic heavy quark action. $N_f$
  is the number of dynamical light quark flavors used.
  \label{table:light}}
\end{table}

\section{Conclusion and Outlook}
Our current understanding of hadronic states containing excited glue
is poor, but recent lattice simulations have shed some light on their
properties.  The glueball spectrum in the pure gauge theory is now well
known, and pioneering studies of the mixings of the scalar glueball
with scalar quarkonia suggests that these mixings may actually be small.
The validity of a Born-Oppenheimer treatment for heavy-quark mesons,
both conventional and hybrid, has been verified at leading order in the
absence of light sea quark effects, and quark spin interactions do
not seem to spoil this.  Progress in including the light sea quarks 
is also being made, and it seems likely that a handful of heavy-quark
hybrid states might survive their inclusion.  Of course, much more work
is needed.  The inner structure of glueballs and flux tubes will be
probed.  Future lattice simulations should provide insight into
hybrid meson production and decay mechanisms and the spectrum and
nature of hybrid baryons; virtually nothing is known about either of
these topics.  Glueballs, hybrid mesons, and hybrid baryons, remain a
potentially rich source of information (and perhaps surprises) about
the confining properties of QCD.  This work was supported by the 
U.S.\ National Science Foundation under award PHY-0099450.

\end{document}